\documentclass[conference]{IEEEtran}
\IEEEoverridecommandlockouts

\usepackage{cite}
\usepackage{amsmath,amssymb,amsfonts}
\usepackage{algorithmic}
\usepackage{graphicx}
\usepackage{textcomp}
\usepackage{xcolor}
\usepackage{tikz}
\usepackage{algorithm}
\usepackage{siunitx}
\usepackage{subcaption}
\usepackage{hyperref}

\def\BibTeX{{\rm B\kern-.05em{\sc i\kern-.025em b}\kern-.08em
    T\kern-.1667em\lower.7ex\hbox{E}\kern-.125emX}}

\begin{document}

\title{GNSS Array-Based Multipath Detection Employing UKF on Manifolds
\thanks{This work was supported by the King Abdullah University of Science and Technology (KAUST) Office of Sponsored Research (OSR) under Award ORA-CRG2021-4695.}}


\author{\IEEEauthorblockN{Abdelgabar Ahmed, Tarig Ballal, Xing Liu, Mohanad Ahmed, and Tareq Y. Al-Naffouri}
\IEEEauthorblockA{\textit{Computer, Electrical and Mathematical Science and Engineering Division} \\
\textit{King Abdullah University of Science and Technology (KAUST)}\\
Thuwal, Saudi Arabia \\
\{\href{mailto:abdelgabar.ahmed@kaust.edu.sa}{abdelgabar.ahmed},
\href{mailto:tarig.ahmed@kaust.edu.sa}{tarig.ahmed},
\href{mailto:xing.liu@kaust.edu.sa}{xing.liu},
\href{mailto:mohanad.ahmed@kaust.edu.sa}{mohanad.ahmed},
\href{mailto:tareq.alnaffouri@kaust.edu.sa}{tareq.alnaffouri}\}\href{mailto:abdelgabar.ahmed@kaust.edu.sa}{@kaust.edu.sa}
}}

\maketitle

\begin{abstract}
Global Navigation Satellite Systems (GNSS) applications are often hindered by various sources of error, with multipath interference being one of the most challenging, particularly in urban environments. In this work, we build on previous research by implementing a GNSS array-based multipath detection algorithm, incorporating real-time attitude estimation for dynamic scenarios. The method fuses GNSS and IMU data using an Unscented Kalman Filter (UKF) on a manifold, enabling continuous attitude tracking. The proposed approach utilizes attitude information from satellite combinations to identify and exclude multipath-affected satellites, improving the accuracy of both positioning and attitude determination. To address computational challenges associated with evaluating large numbers of satellite combinations, we propose the use of the Random Sample Consensus (RANSAC) algorithm, which reduces the number of combinations assessed while maintaining high detection performance. Performance evaluations are conducted using trajectories and IMU readings from the KITTI dataset. GNSS observations are simulated based on ground truth positions and satellite ephemeris. The results demonstrate the effectiveness of the proposed approach in detecting satellites affected by multipath interference. Significant improvements in positioning accuracy are observed, particularly in scenarios where a large portion of the visible satellites are contaminated by severe multipath.

\end{abstract}

\section{Introduction}
Multipath is a significant source of error in Global Navigation Satellite Systems (GNSS) applications, particularly in urban environments where it is most prevalent. While other error sources, such as ionospheric and tropospheric disturbances and clock inaccuracies, can often be mitigated using well-established techniques, addressing the impact of multipath remains a persistent challenge. In dense urban areas, tall buildings and other reflective surfaces create multiple GNSS signal paths to the receiver. These reflected signals introduce errors in pseudorange and carrier-phase measurements, severely degrading positioning accuracy \cite{chengyan2014multipath}.

In urban intelligent transportation systems, GNSS remains indispensable for absolute positioning in autonomous vehicles, despite advancements in onboard sensor fusion. Multipath effects—caused by signal reflections from buildings and infrastructure—persist as a critical barrier to reliable positioning accuracy~\cite{liu2019gnss}. While sensor-aided navigation (e.g., LiDAR, cameras) compensates for GNSS errors, the foundational role of GNSS necessitates robust multipath detection and mitigation strategies to ensure safety-critical performance.

To tackle this challenge, various multipath detection techniques have been explored across different stages of GNSS signal processing, ranging from correlator-level analysis to observable-based classification. At the correlator stage, machine learning techniques such as convolutional neural networks (CNNs) have been applied to raw correlator outputs to distinguish multipath-affected signals \cite{suzuki2020nlos}. Supervised classifiers utilizing features like signal strength, elevation angle, Doppler shift, and pseudorange residuals have also demonstrated success in LOS/NLOS classification \cite{koiloth2024ml}. However, these approaches rely heavily on labeled datasets, which are often challenging to acquire due to the need for accurate ground truth in dynamic urban environments. Furthermore, models trained on specific locations may struggle with generalization, as multipath characteristics vary significantly across different cities, signal conditions, and receiver hardware.

To enhance multipath resilience, recent studies have explored integrating GNSS with complementary sensors such as sky-view cameras and 3D LiDARs, which provide independent environmental measurements for detecting and even correcting NLOS signals \cite{wen20233d, bai2019real}. Another approach leverages 3D city models to mitigate multipath by simulating GNSS signal propagation using ray-tracing techniques \cite{zhang2018gnss}. Additionally, methods such as likelihood-based ranging and shadow matching have been employed in 3D-mapping-aided (3DMA) positioning, improving accuracy by refining satellite visibility predictions based on urban structures \cite{zhong2023optimizing, zhong2022outlier}. While promising, these methods rely on high-fidelity environmental models and face sensor-specific limitations, such as extreme illumination conditions affecting cameras and field-of-view constraints for LiDARs, which can reduce their effectiveness in complex urban environments.

Various methods utilizing GNSS antenna array processing have been explored in the literature at different stages of the GNSS receiver pipeline. For instance, at the signal processing stage, as seen in \cite{zorn2020accurate}, and in array-based observables, as in \cite{chen2021performance}, which assesses baseline consistency between observed and predicted carrier-phase differentials to detect contaminated signals in real time. Additionally, \cite{esswein2025classification} uses GNSS array DoA and pseudorange data to classify spoofed signals via multi-hypothesis tests. 
Array-based approaches, which inherently require multi-antenna setups, can offer reliable mitigation without relying on external sensors. To ensure unambiguous phase observations, the array elements must be spaced by up to a half-wavelength \cite{zorn2020accurate}. The method introduced in \cite{ahmed2023multipath} utilizes carrier-phase observations from a GNSS antenna array for attitude estimation. By leveraging the specific geometry of the antenna array, maintaining synchronization across GNSS receivers, and incorporating prior information about the array's attitude, this approach aims to identify satellite observations affected by multipath contamination.

In this work, we extend the concept of multipath detection using GNSS antenna arrays and attitude validation, as introduced in \cite{ahmed2023multipath}, to real-time scenarios. By incorporating a filter that fuses GNSS and IMU data, we enable continuous attitude estimation, which is then used in each iteration to identify and exclude multipath-affected satellites. Additionally, we propose a computationally efficient algorithm to distinguish between multipath-contaminated and multipath-free satellite observations by examining various satellite combinations. This novel approach utilizes the Random Sample Consensus (RANSAC) algorithm \cite{fischler1981random}, which significantly reduces the number of combinations that need to be evaluated while maintaining high multipath detection performance.

Filtering-based approaches, particularly the Unscented Kalman Filter (UKF), have been widely adopted for navigation applications to fuse GNSS with measurements from sensors such as  IMUs \cite{alaba2024gps}, due to their ability to handle nonlinear system dynamics. Recent advancements have integrated UKF with manifold optimization to account for the geometric constraints of pose data. For instance, \cite{cantelobre2020real, brossard2020code} implemented an UKF on the SO(3) manifold for navigation, estimating both attitude and location. In our implementation, we utilize UKF on manifold to consider the geometric constraint for the pose data, including attitude, and position.

The rest of the paper is structured as follows. In Section \ref{Problem}, we describe the problem and provide background material on the main challenges in multipath detection. In Section \ref{Method}, we present our proposed solution, including the filter for real-time attitude determination and the computationally efficient algorithm for multipath detection. Section \ref{Sim} showcases the simulation setup and results, demonstrating the effectiveness of our approach in realistic, dynamically simulated scenarios. Finally, Section \ref{Concl} concludes the paper and highlights the main future directions.

\section{Problem Description and Background} \label{Problem}
\subsection{Problem Description}
\label{subsec:problem:description}
\begin{figure}[htbp] 
    \centering
    \includegraphics[width=0.4\textwidth]{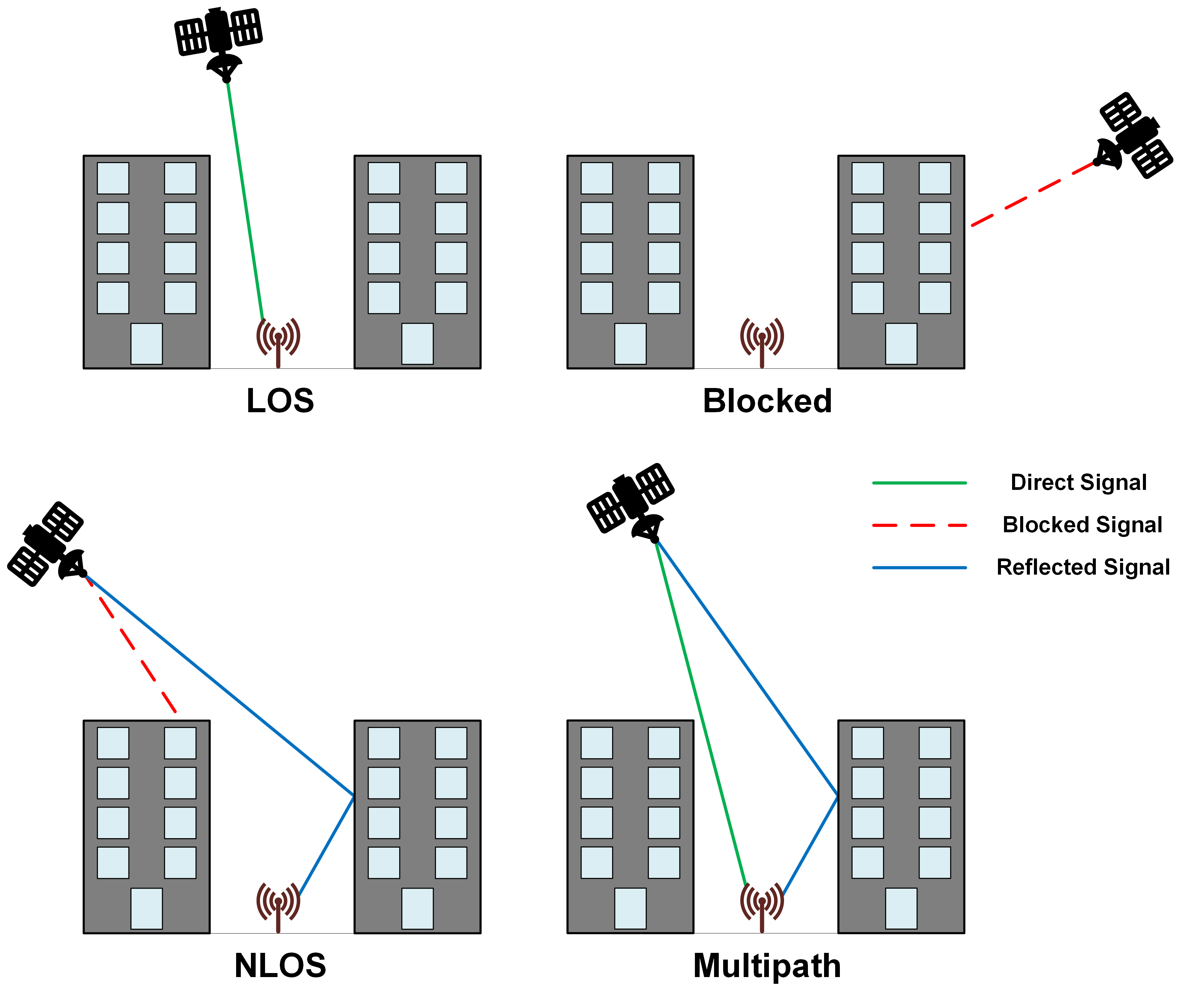} 
    \caption{Signal Reception Scenarios for GNSS Receivers in Urban Environments}
    \label{fig:rec_sc}
\end{figure}

The reception scenarios for a GNSS receiver from a satellite \( s \), as illustrated in Fig.~\ref{fig:rec_sc}, depend on the satellite's elevation and the geometry of the surrounding environment. Depending on these factors, the received signal can either follow a direct path to the receiver, reflect off surrounding surfaces resulting in indirect signal paths, or experience a combination of both direct and reflected signals. In some cases, the signal may be completely blocked, preventing any reception. These scenarios can be characterized by the direct signal amplitude \( a_0^s \) and the amplitudes of the indirect paths \( \{a_i^s\}_{i=1}^{N_{MP}^s} \), where \( N_{MP}^s \) represents the number of indirect paths from satellite \( s \) to the receiver. These scenarios are summarized as follows \cite{ahmed2023multipath}:

\begin{itemize}
    \item \textbf{LOS (Line-of-Sight):} Only the direct path is present. This occurs when \( a_0^s \gg 0 \) and \( a_i^s \approx 0 \) for all \( i > 0 \).
    \item \textbf{NLOS (Non-Line-of-Sight):} Only an indirect path is present. This occurs when \( a_0^s \approx 0 \) and \( a_i^s \gg 0 \) for any \( i \in\{1, \cdots, N_{MP}^s \} \).
    \item \textbf{Multipath:} Both direct and indirect paths are present. This occurs when \( a_0^s \gg 0 \) and \( a_i^s \gg 0 \) for any \( i \in\{1, \cdots, N_{MP}^s \} \).
    \item \textbf{Blocked:} No signal paths are present. This occurs when \( a_0^s \approx 0 \) and \( a_i^s \approx 0 \) for all \( i \in\{1, \cdots, N_{MP}^s \} \).
\end{itemize}

The problem we aim to address involves a vehicle navigating through an urban area, where tall obstructions block or reflect satellite signals, leading to NLOS and multipath signal reception. In such cases, sizable errors may occur in the GNSS observations, which can significantly degrade the accuracy of the positioning solution. Identifying GNSS observations with large errors due to NLOS and multipath reception is very critical for 
 maintaining the fidelity of the positioning solution by rejecting those contaminated observations. As such, the objective herein is to determine, for each satellite observation, whether it is \emph{clean} (LOS) or \emph{corrupted} (either NLOS or multipath). Only signals unaffected by NLOS and multipath are then utilized for positioning. This detection and mitigation process should be performed in real-time as the vehicle moves.

\subsection{Related Work}
\label{subsec:problem:related}
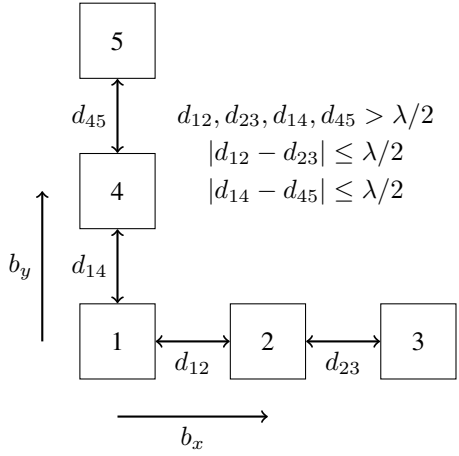
\begin{figure}[H]
    \centering
    \begin{tikzpicture}
    \node[draw, minimum size=1cm] (1) at (0, 0) {1};
    \node[draw, minimum size=1cm] (2) at (2, 0) {2};
    \node[draw, minimum size=1cm] (3) at (4, 0) {3};
    \node[draw, minimum size=1cm] (4) at (0, 2) {4};
    \node[draw, minimum size=1cm] (5) at (0, 4) {5};

    \draw[<->, thick] (1) -- (2) node[midway, below] {$d_{12}$};
    \draw[<->, thick] (2) -- (3) node[midway, below] {$d_{23}$};
    \draw[<->, thick] (1) -- (4) node[midway, left] {$d_{14}$};
    \draw[<->, thick] (4) -- (5) node[midway, left] {$d_{45}$};

    \node at (2.5, 3) {$d_{12}, d_{23}, d_{14}, d_{45} > \lambda/2$};
    \node at (2.5, 2.5) {$|d_{12} - d_{23}| \leq \lambda/2$};
    \node at (2.5, 2) {$|d_{14} - d_{45}| \leq \lambda/2$};

    \draw[->, thick] (0, -1) -- (2, -1) node[midway, below] {$b_x$};
    \draw[->, thick] (-1, 0) -- (-1, 2) node[midway, left] {$b_y$};
    \end{tikzpicture}
    \caption{The proposed antenna array configuration with 5 elements. Inter-element distances and baseline direction vectors are indicated.}
    \label{fig:array}
\end{figure}

Building on the antenna array geometry of five synchronized receivers (or antennas) proposed in \cite{ahmed2023multipath}, our approach employs an Unscented Kalman Filter (UKF) to fuse the GNSS array observations with an IMU. The UKF tracks the vehicle's attitude, enabling the identification of multipath-contaminated satellite observations by detecting inconsistencies between the attitude estimates derived from satellite observations and those estimated by the filter. 
Since the attitude determination process will run repeatedly in real-time, the algorithm proposed in \cite{ballal2014gnss,douik2019precise,liuGnss2018} offers a feasible approach due to its computational simplicity and suitability for scenarios with a low number of satellites. This is particularly relevant when assessing various satellite combinations to identify those affected by multipath.
Fig. (\ref{fig:array}) depicts the antenna array configuration applied to tackle the multipath problem. Due to the relatively short distances between the antennas compared to the array distance from street-side reflectors, it can be reasonably assumed that all antennas experience the same reception scenario simultaneously. Both direct and reflected signals received by the array undergo the same reflection conditions, resulting in similar amplitudes. As such, the primary difference between the signals lies in their phase delay, which arises due to variations in the excess path length corresponding to the inter-receiver distances.  

For the array in Fig. (\ref{fig:array}), a sufficient condition for phase-difference disambiguation is that the baselines satisfy $|d_{12}-d_{23}| \leq \lambda/2$ and $|d_{14}-d_{45} \leq \lambda/2|$, where $\lambda$ is the wavelength of the operational frequency \cite{ballal2014gnss}. Under these conditions, it is shown that the ambiguity resolution problem associated with the attitude determination process can be solved in a computationally simplified manner. Our proposed approach to multipath detection capitalizes on this computational efficiency to deliver a practical solution.

\subsection{GNSS Observation Models in Multipath Scenarios}
\label{subsec:problem:models}

We assume that the received GNSS signals at each antenna combine additively. Thus, the carrier phase observable error due to multipath contamination for reception from satellite \( s \) at antenna \( r \) can be modeled by summing the error contributions from \( N_{MP}^s \) multipath signals. Each multipath signal \( k \in \{1, \cdots, N_{MP}^s\} \) has an amplitude \( a_k^s \) and a carrier phase delay \( \Delta\phi_{k,r}^s \) due to its indirect arrival. For antenna \( r \), the carrier phase error is given by \cite{fan2006estimation, smyrnaios2013multipath}:

\begin{equation}
\psi_r^s = \arctan\left(\frac{\displaystyle\sum_{k=1}^{N_{MP}^s} a_k^s \sin\left(\Delta\phi_{k,r}^s\right)}{1 + \displaystyle\sum_{k=1}^{N_{MP}^s} a_k^s \cos\left(\Delta\phi_{k,r}^s\right)}\right),
\label{eq:carrier_phase_error}
\end{equation}

The resulting composite carrier phase measurement at antenna \( r \) is given by:

\begin{equation}
\bar{\phi}_r^s = \phi_r^s + \psi_r^s,
\label{eq:composite_phase_other}
\end{equation}
where \( \phi_r^s \) is the direct signal's carrier phase and \( \psi_r^s \) is the induced multipath error at antenna \( r \).

Similarly, the pseudorange error at antenna \( r \) is given by \cite{smyrnaios2013multipath}:

\begin{equation}
\Delta p_r^s = \frac{\displaystyle\sum_{k=1}^{N_{MP}^s} a_k^s \delta_{k,r}^s \sin\left(\Delta\phi_{k,r}^s\right)}{1 + \displaystyle\sum_{k=1}^{N_{MP}^s} a_k^s \cos\left(\Delta\phi_{k,r}^s\right)},
\label{eq:pseudorange_error}
\end{equation}
where \( \delta_{k,r}^s \) is the excess path length of the reflected signal. For antennas \( r \neq 1 \), the excess path length is related to that of antenna 1 according to the relative geometry between antennas:

\begin{equation}
\delta_{k,r}^s = \delta_{k,1}^s - \mathbf{d}_{1r} \cdot \mathbf{q}_k^s,
\label{eq:path_length_other}
\end{equation}
where \( \mathbf{q}_k^s \) is the unit vector representing the direction of arrival of signal \( k \) from satellite \( s \), and \( \mathbf{d}_{1r} \) is the vector from antenna 1 to antenna \( r \). The excess path length \( \delta_{k,r}^s \) corresponding to a reflection point \( \mathbf{o} \) in the environment is calculated as:

\begin{equation}
\delta_{k,r}^s = \|\mathbf{x}_s - \mathbf{o}\| + \|\mathbf{o} - \mathbf{x}_r\| - \|\mathbf{x}_s - \mathbf{x}_r\|
\label{eq:excess_path_length}
\end{equation}
where \( \mathbf{x}_s \), \( \mathbf{x}_1 \), and \( \mathbf{o} \) are the positions of the satellite, antenna \( r \), and reflection point, respectively. This gives the additional distance traveled by the reflected signal relative to the direct path.
Once the excess path length \( \delta_{k,r}^s \) is obtained, the corresponding carrier phase delay \( \Delta\phi_{k,r}^s \) is calculated as:
\begin{equation}
\Delta\phi_{k,r}^s = \frac{2\pi}{\lambda} \, \delta_{k,r}^s.
\label{eq:carrierphase_delay}
\end{equation}

To account for phase ambiguity, the resulting delay is wrapped within one full phase cycle, i.e., to the interval \( (-\pi, \pi) \).
Finally, the contaminated pseudorange measurement at antenna \( r \) is obtained by summing the direct pseudorange \( p_r^s \) and the error:

\begin{equation}
\bar{p}_r^s = p_r^s + \Delta p_r^s.
\label{eq:contaminated_pseudorange}
\end{equation}

These models provide a detailed representation of how multipath effects are modeled on both carrier phase and pseudorange measurements received at array antennas. They form the basis for the realistic GNSS multipath scenarios used in Section~\ref{Sim} to validate the performance of the proposed multipath detection method.

\section{The proposed multipath detection and mitigation method} \label{Method}

\begin{figure}[htbp] 
    \centering
    \includegraphics[width=0.5\textwidth]{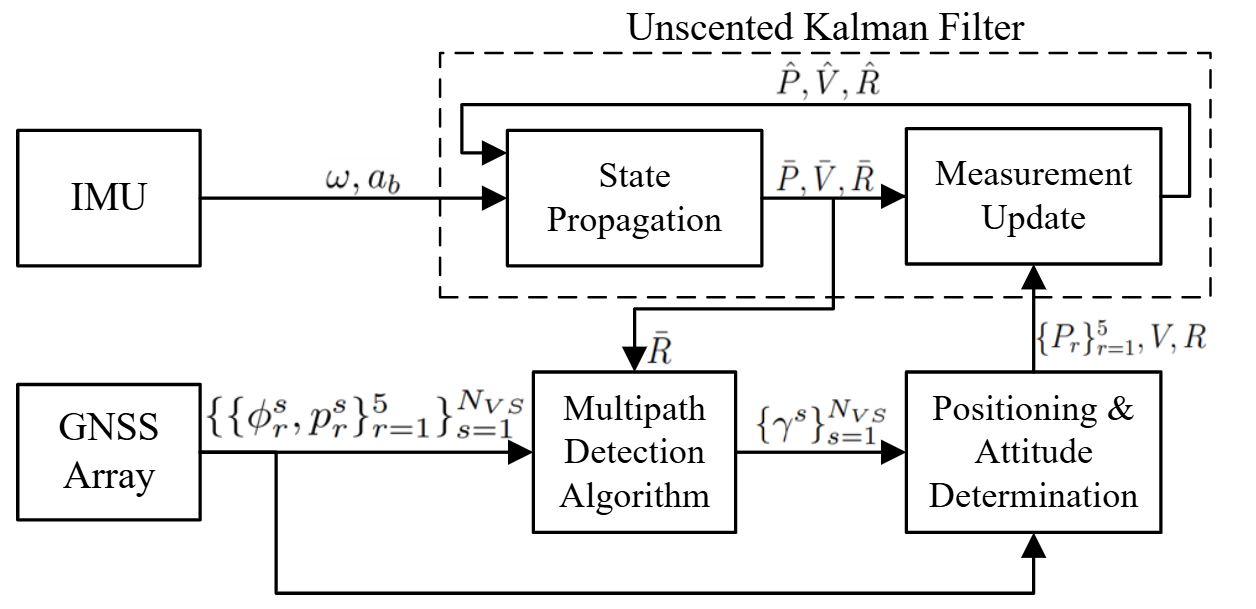} 
    \caption{GNSS Array-IMU Fusion Framework for Multipath Detection and Positioning}
    \label{fig:UKF_Block}
\end{figure}

Fig.~(\ref{fig:UKF_Block}) illustrates the proposed framework for dynamic multipath detection and refined GNSS-based positioning and attitude estimation. The Unscented Kalman Filter (UKF) estimates the vehicle’s state—position (\(P\)) of receiver 1 in the array, velocity (\(V\)), and array attitude (\(R\))—by integrating inertial measurements from the IMU and GNSS observations from a multi-antenna array, as configured in Fig.~(\ref{fig:array}).
The IMU supplies angular velocity (\(\omega\)) and acceleration (\(a_b\)), which are used during the UKF state propagation step. GNSS observations consist of pseudorange (\(p_r^s\)) and carrier phase (\(\phi_r^s\)) measurements for each antenna \(r = 1,\dots,5\) and satellite \(s = 1,\dots,N_{VS}\), where \(N_{VS}\) denotes the number of visible satellites.
After state propagation, the predicted attitude (\(\bar{R}_n\)) is extracted and provided to the multipath detection algorithm. This algorithm assesses the geometric consistency of the incoming satellite signals, enabling the identification and exclusion of multipath-affected signals. It yields a refined set of satellite indicators \(\{\gamma^s\}_{s=1}^{N_{VS}}\), where \(\gamma^s = 1\) denotes a multipath-contaminated satellite and \(\gamma^s = 0\) indicates a multipath-free satellite.
Only the multipath-free satellites are used in the positioning and attitude determination module. The GNSS array attitude estimation is computed using carrier-phase observations, while the position solutions for each of the five GNSS receivers are separately obtained as independent Single Point Positioning (SPP) solutions using pseudorange observations. While incorporating carrier-phase processing between multiple receivers could further improve the GNSS array positioning solution, the focus of this implementation remains on pseudorange-based SPP to emphasize the contributions of the filter and multipath detection algorithm in refining the solution.

\subsection{GNSS-IMU Fusion via UKF on Manifolds}

We propose a UKF-based approach~\cite{wan2000unscented} for estimating the position, velocity, and attitude of a GNSS antenna array by integrating inertial measurements from an IMU with GNSS observations, while accounting for the relative configuration among the receivers in the array. The UKF operates on parallelizable manifolds~\cite{boumal2023introduction}, following the implementation in~\cite{brossard2020code}, with the system state embedded in the space $SO(3) \times \mathbb{R}^{12}$.

\subsubsection{State Representation}
The system state is defined as:
\begin{itemize}
    \item \textbf{3D attitude:} Rotation matrix $R \in SO(3)$.
    \item \textbf{3D position:} $P \in \mathbb{R}^3$ (ENU coordinates).
    \item \textbf{Velocity:} $V \in \mathbb{R}^3$.
    \item \textbf{IMU gyroscope bias:} $b_g \in \mathbb{R}^3$.
    \item \textbf{IMU accelerometer bias:} $b_a \in \mathbb{R}^3$.
\end{itemize}

The filter inputs for state propagation are:
\begin{itemize}
    \item \textbf{Gyroscope measurements:} $\omega \in \mathbb{R}^3$.
    \item \textbf{Accelerometer measurements:} $a_b \in \mathbb{R}^3$ (body frame).

\end{itemize}

The filter provides the following outputs:

\begin{itemize}
    \item \textbf{Array attitude:} Euler angles \( \theta \in \mathbb{R}^3 \) (roll, pitch, yaw).
    \item \textbf{Position of the 5 receivers:} $\{ P_r \}_{r=1}^5 \in \mathbb{R}^3$.
    \item \textbf{Velocity of the vehicle:} $V \in \mathbb{R}^3$.
\end{itemize}

\subsubsection{State Propagation}

The state propagation from time step \( n \) to \( n+1 \) over a time interval \( \Delta t \) is expressed as:
\begin{equation}
\begin{aligned}
    \bar{R}_{n+1} &= \hat{R}_n \exp\left( (\omega - \hat{b}_{g,n}) \Delta t \right), \\
    \bar{P}_{n+1} &= \hat{P}_n + \hat{V}_n \Delta t + \frac{1}{2} a (\Delta t)^2, \\
    \bar{V}_{n+1} &= \hat{V}_n + a \Delta t, \\
    \bar{b}_{g,n+1} &= \hat{b}_{g,n} + w_{g,n} \Delta t, \\
    \bar{b}_{a,n+1} &= \hat{b}_{a,n} + w_{a,n} \Delta t,
\end{aligned}
\label{eq:UKF_Propagate}
\end{equation}
where \( \hat{X}_n \) denotes the estimated state from the previous filter update, and \( \bar{X}_n \) denotes the predicted state after propagation through the system model. Here \( w_{g,n} \) and \( w_{a,n} \) are random walk noise terms affecting the gyroscope and accelerometer biases, respectively. Both biases, \( {b}_g \) and \( {b}_a \), are modeled as random walk processes.
The acceleration in the global frame \( a \) is given by:
\begin{equation}
    a = \hat{R}_n (a_b - \hat{b}_{a,n}) + g,
    \label{eq:Global_Acceleration}
\end{equation} 
where \( g \) denotes the gravity vector.

\subsubsection{Observation Model}

After excluding multipath-affected satellites, the remaining GNSS observations are used to extract the receiver positions \( \{ P_r \}_{r=1}^5 \), and array attitude (\( R \)). The Euler angles are then obtained by converting the rotation matrix \( R \) into roll, pitch, and yaw, following \( \theta = \text{SO3\_to\_RPY}(R) \).
The predicted outputs from the UKF state propagation are:
\begin{equation}
\begin{aligned}
    &\bar{\theta}_n = SO3\_to\_RPY(\bar{R}_n), \\
    &\bar{P}_{1,n} = \bar{P}_n, \\
    &\bar{P}_{r,n} = \bar{P}_n + d_{1r} \bar{R}_n \begin{bmatrix}1 \\ 0 \\ 0\end{bmatrix}, \quad r = 2,3, \\
    &\bar{P}_{r,n} = \bar{P}_n + d_{1r} \bar{R}_n \begin{bmatrix}0 \\ 1 \\ 0\end{bmatrix}, \quad r = 4,5.
\end{aligned}
\label{eq:UKF Obs}
\end{equation}
here, \(r = 2, 3\) corresponds to antennas on the x-axis, and \(r = 4, 5\) corresponds to antennas on the y-axis, as illustrated in Fig.~(\ref{fig:array}). The position predictions for \(r = 2, 3\) are based on the first column of the attitude matrix \(\bar{R}_n\), which represents the x-axis baseline in the rotated coordinate system. For \(r = 4, 5\), the second column of \(\bar{R}_n\) is used, which corresponds to the y-axis baseline in the rotated coordinate system.

The predicted outputs in Eq. \eqref{eq:UKF Obs} are then used to compute the residual relative to the array-derived positions and attitude solutions (\( \{ P_r \}_{r=1}^5 , R\)). This residual is then used to update the UKF state and covariance through the Kalman gain. The filter operates on a manifold, with retraction and inverse retraction defined using the $SO(3)$ exponential and logarithm for the attitude component, and standard vector operations for the remaining states, following the implementation described in~\cite{brossard2020code}.

\subsection{Multipath Detection Algorithm}

The core principle of this approach is that attitude estimation is based on a collective set of satellite measurements. By selecting only the satellites unaffected by multipath, more accurate attitude estimates can be achieved compared to using all available satellites, including those contaminated by multipath. The attitude is estimated as outlined in \cite{ahmed2023multipath, ballal2014gnss}, where the same receiver configuration depicted in Fig.~(\ref{fig:array}) was used to compute the attitude matrix. This is done using carrier phase observations from the five receivers and the line-of-sight (LOS) matrix \(H\). By tracking the positions of the receivers within the implemented filter, the LOS vectors from each satellite to the receivers can be determined. The attitude estimation algorithm, as detailed in \cite{ahmed2023multipath, ballal2014gnss}, is integrated into our proposed method in Algorithm~(\ref{alg:multipath_detection}), referred to as \textit{AttFromPhase}. This function estimates the attitude for a set of satellites using carrier phase observations from the five receivers, the LOS matrix \(H\), and the inter-distances between the antennas.


In \cite{ahmed2023multipath}, the discrimination algorithm evaluates all possible subsets of the visible satellites \( N_{VS} \) that contain at least \( N_{Smin} \) satellites. The attitude estimated from each subset is compared to the reference attitude, with the deviation measured as the geodesic distance on \( SO(3) \) \cite{absil2008optimization}. These subsets are then clustered using the DBSCAN algorithm \cite{ester1996density}, and the free satellites are selected from the largest cluster with the minimum error. In urban canyons, the number of visible satellites \( N_{VS} \) is typically small, making the number of combinations manageable. However, as \( N_{VS} \) increases, the number of possible combinations grows rapidly, becoming much larger (over 32K for \( N_{VS} = 15 \) and \( N_{Smin} = 4 \)) and requiring significant computational resources. To address this challenge, we propose an alternative approach that reduces the number of satellite combinations assessed, while maintaining competitive multipath discrimination performance.

The proposed algorithm is based on the outlier detection method known as RANSAC \cite{fischler1981random}, which is commonly used in fields such as image processing for feature matching. This technique has also been successfully adapted for various navigation applications, as shown in\cite{jiang2012gnss} and \cite{zhang2023ransac}.

\begin{table}[H]
    \centering
    \begin{tabular}{ll}
        $H \in \mathbb{R}^{N_{SV} \times 3}$ & Satellite LOS matrix from estimated position \\ &  and GPS constellation. \\
        $\phi_{1:5} \in \mathbb{R}^{N_{SV}}$ & Carrier phase observations at the 5 receivers. \\
        $R_{ref} \in SO(3)$ & Reference attitude prediction from UKF (\( \bar{R}_n \)). \\
        $d_{12}, d_{23}, d_{14}, d_{45}$ & Inter-receiver distances (m). \\
        $\epsilon_{inlier}$ & Attitude error threshold for inlier acceptance. \\
        $N_{min}$ & Minimum number of inliers required for a valid solution. \\
        $N_{iter}$ & Maximum RANSAC iterations, Eq. (\ref{eq:ransac_iterations}). \\
        $m \geq 3$ & Minimum number of satellites required for \\ & attitude estimation. \\
        \hline
    \end{tabular}
    \caption{Inputs used in Algorithm~(\ref{alg:multipath_detection})}
\end{table}

\begin{algorithm}[H] 
\caption{GNSS Array-Based Multipath Detection}\label{alg:multipath_detection}
\begin{algorithmic}[1]
\STATE \textbf{Initialize:}
\STATE $R_{best} \gets \text{AttFromPhase}(\phi, d, H)$ 
\STATE $S_{best} \gets \{1,\ldots,N_{VS}\}$ 
\STATE $er_{best} \gets d_g(R_{ref}, R_{best})$ 
\COMMENT{Geodesic distance on $SO(3)$}\\
\STATE $\gamma \gets \mathbf{1}_{N_{VS}}$ \COMMENT{Initialize all $\gamma^s = 1$}

\STATE \textbf{Generate combinations:}
\STATE $Sat\_Combs \gets \{ S_1, S_2, ..., S_{ \binom{N_{VS}}{m} } \}$ 

\STATE \textbf{Main loop (RANSAC):}
\FOR{$iter = 1$ \TO $N_{iter}$}
    \STATE Randomly select $S_k \subset Sat\_Combs$ 
    \\
    \COMMENT{Hypothesized clean set}
    \STATE $I_k \gets S_k$ \COMMENT{Initialize inliers set with seed satellites}
    \STATE $R_k \gets \text{AttFromPhase}(\phi(S_k), d, H(S_k))$
    
    \STATE \textbf{Inlier expansion:}
    \FORALL{$sv \in \{1,\ldots,N_{VS}\} \setminus S_k$}
        \STATE $R_{sv} \gets \text{AttFromPhase}( \phi(S_k \cup sv), d, H(S_k \cup sv) )$
        \STATE $er_{sv} \gets d_g(R_k, R_{sv})$ \COMMENT{Consistency check}
        \IF{$er_{sv} < \epsilon_{inlier}$}
            \STATE $I_k \gets I_k \cup \{sv\}$
        \ENDIF
    \ENDFOR
    
    \IF{$|I_k| \geq N_{min}$}
        \STATE $R_k \gets \text{AttFromPhase}(\phi(I_k), d, H(I_k))$
        \STATE $er_k \gets d_g(R_{ref}, R_k)$
        \IF{$er_k < er_{best}$}
            \STATE $R_{best}, S_{best}, er_{best} \gets R_k, I_k, er_k$
        \ENDIF
    \ENDIF
\ENDFOR

\STATE \textbf{Assign $\gamma$:}
\FOR{$s = 1$ \TO $N_{VS}$}
    \IF{$s \in S_{best}$}
        \STATE $\gamma^s \gets 0$ \COMMENT{multipath-free}
    \ELSE
        \STATE $\gamma^s \gets 1$ \COMMENT{multipath-contaminated}
    \ENDIF
\ENDFOR

\STATE \textbf{Return:} $\{\gamma\}_{s=1}^{N_{VS}}$, $R_{best}$, $er_{best}$
\end{algorithmic}
\end{algorithm}

\begin{figure*}[ht]
    \centering
    \includegraphics[width=0.9\textwidth]{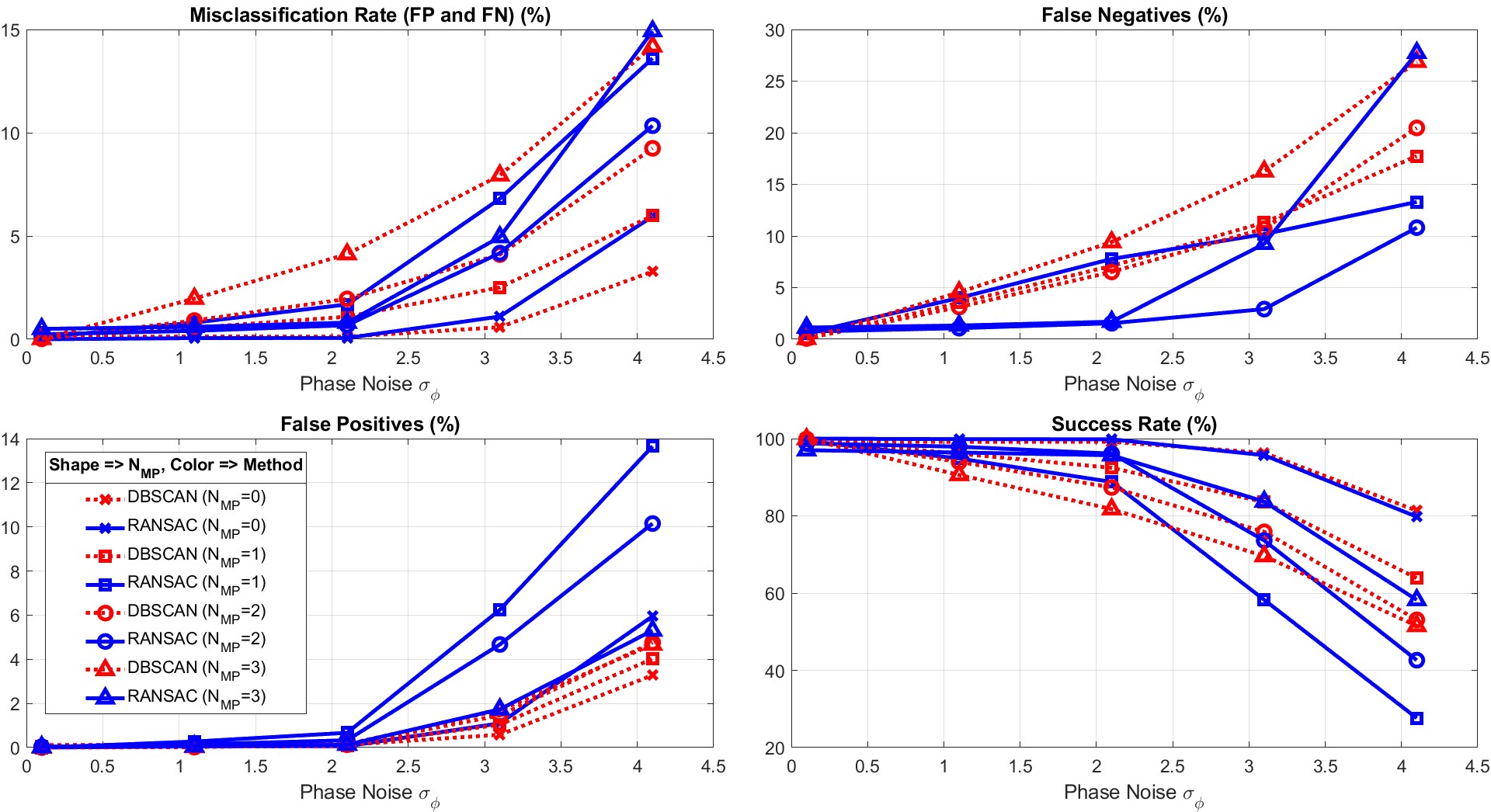}
    \caption{Classification performance comparison between DBSCAN (red) and proposed method (blue) across phase noise levels.}
    \label{fig:success_rate}
\end{figure*}

 Our approach incorporates a RANSAC-inspired framework for multipath detection, starting with the assumption that all \( N_{VS} \) visible satellites are free from multipath errors. Using all available measurements, we calculate an initial attitude estimate. The core of the algorithm follows a hypothesize-and-verify paradigm. In each RANSAC iteration, a subset of \( m \) satellites is randomly selected as potential inliers, and the array attitude is estimated using only this subset. The inlier set is then expanded by checking the consistency of additional satellites with the current estimate. If the candidate set satisfies the geometric constraints, it is designated as a potential solution. 

A satellite \( sv \) is included in the inlier set if its addition results in minimal attitude perturbation, ensuring geometric consistency with the hypothesized clean set. The algorithm then selects a sufficiently large consensus set that minimizes the geodesic distance to the UKF-predicted attitude \( \bar{R}_n \), which is used as \( R_{\text{ref}} \) in the algorithm. The final error \( er_{best} \) quantifies estimation uncertainty and is used in subsequent filtering within the UKF.

The geodesic distance \( d_g(\cdot,\cdot) \) on \( SO(3) \) is computed using:
\begin{equation}
    d_g(R_1,R_2) = \|\log(R_1^\top R_2)\|_F, \label{eq:geodesic_distance}
\end{equation}
where \( \log(\cdot) \) represents the matrix logarithm. 

The number of iterations \( N_{iter} \) is determined using RANSAC’s probabilistic model \cite{fischler1981random}:  
\begin{equation}
    N_{iter} = \frac{\log(1-p)}{\log(1-(1-\eta)^m)}, \label{eq:ransac_iterations}
\end{equation}
where \( p \) is the desired success probability for selecting a clean set, \( \eta \) denotes the assumed ratio of outlier satellites contaminated by multipath, and \( m \) is the minimum number of satellites needed to construct the attitude estimation model. The implementation of the proposed multipath detection method is presented in Algorithm~(\ref{alg:multipath_detection}).

\section{Performance Evaluation} \label{Sim}

\begin{figure*}[ht]
    \centering
    \includegraphics[width=0.9\linewidth]{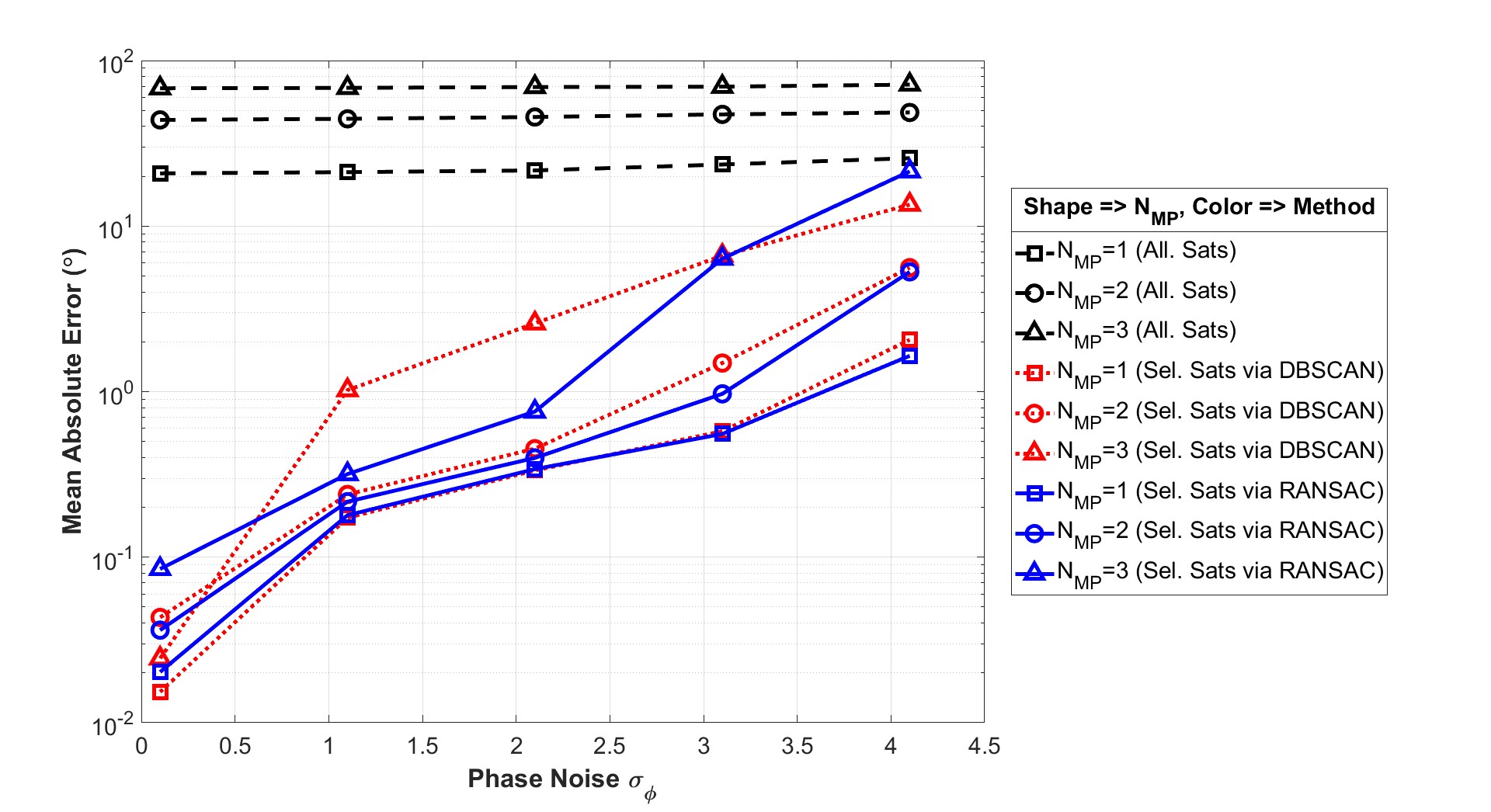}
    \caption{Baseline MAE comparison. Proposed method outperforms DBSCAN in attitude estimation accuracy.}
    \label{fig:baseline_mae}
\end{figure*}

\subsection{Experimental Setup} 
Two simulation scenarios were designed to assess the performance of the proposed GNSS array-based multipath detection and exclusion approach using algorithm~(\ref{alg:multipath_detection}):

\begin{itemize}
    \item \textbf{Static Receiver Scenario:} Evaluated the method’s multipath detection accuracy against the DBSCAN-based multipath detector introduced in \cite{ahmed2023multipath}.
    \item \textbf{Moving Vehicle Scenario:} Assessed position and attitude estimation performance using real-world vehicle trajectories from the KITTI dataset \cite{geiger2013vision}.
\end{itemize}

In both scenarios, GPS L1 signals (\( \lambda_1 \approx 19.05 \, \text{cm} \)) were employed. Pseudorange and carrier phase measurements were generated based on satellite positions obtained from broadcast ephemeris and the known ground-truth positions along the trajectories. These measurements were then corrupted with additive zero-mean white Gaussian noise (AWGN), characterized by standard deviations \( \sigma_{\rho} \) and \( \sigma_{\phi} \) for the pseudorange and carrier phase, respectively. Multipath effects were simulated according to the models described in Eqs.~\eqref{eq:carrier_phase_error}--\eqref{eq:contaminated_pseudorange}.

\subsection{Simulation 1: Static Multipath Detection Benchmark}

This simulation focused on identifying multipath-contaminated satellite observations. The classification performance of the proposed method was compared against the DBSCAN-based method from \cite{ahmed2023multipath}, using the following metrics:

\begin{enumerate}
    \item \textbf{Success Rate:} Percentage of cases where all satellite observations were correctly classified;
    \item \textbf{False-Negative Rate:} Proportion of multipath-contaminated satellite observations incorrectly labeled as clean;
    \item \textbf{False-Positive Rate:} Proportion of clean satellite observations incorrectly labeled as contaminated;
    \item \textbf{Misclassification Rate:} Overall percentage of satellites misclassified, including both false positives and false negatives;
    \item \textbf{Baseline MAE:} Error in the estimated attitude after excluding detected contaminated satellites.
\end{enumerate}

To compute these metrics, a Monte Carlo simulation approach was used with a large number of independent trials. For each trial, a random subset of \( N_{\text{MP}} \) satellites (out of \( N_{\text{SV}} = 7 \)) was selected to be affected by multipath. For each affected satellite, a reflection point \( o \) was randomly chosen from the surrounding environment, and the corresponding multipath error was generated according to the models described in Eqs.~\eqref{eq:carrier_phase_error}--\eqref{eq:contaminated_pseudorange}. The classification results from each trial were aggregated to compute the statistical rates.
Figures~(\ref{fig:success_rate}) and (\ref{fig:baseline_mae}) illustrate the classification performance and attitude estimation accuracy—represented by the baseline MAE— across varying carrier phase noise levels and different values of \( N_{\text{MP}} \).

Although the DBSCAN method achieves slightly higher overall classification success rates, our proposed algorithm more reliably detects contaminated signals, as indicated by lower false-negative rates. While this conservative strategy increases false positives—sometimes classifying clean signals as contaminated—it results in cleaner observation sets that improve downstream attitude and position estimation.

Figure~(\ref{fig:baseline_mae}) presents the baseline MAE in degrees, showing attitude estimation accuracy under three setups: using all satellites (including multipath-affected), using satellites filtered via the DBSCAN-based method~\cite{ahmed2023multipath}, and using those filtered by our algorithm~(\ref{alg:multipath_detection}). Excluding contaminated satellites clearly enhances accuracy. Despite assessing fewer satellite combinations, our method remains competitive while reducing computational complexity—especially beneficial as the number of observed satellites increases—making it well-suited for real-time applications.

\begin{figure*}[ht]
    \centering
    \begin{subfigure}[b]{0.48\linewidth}
        \centering
        \includegraphics[width=\linewidth]{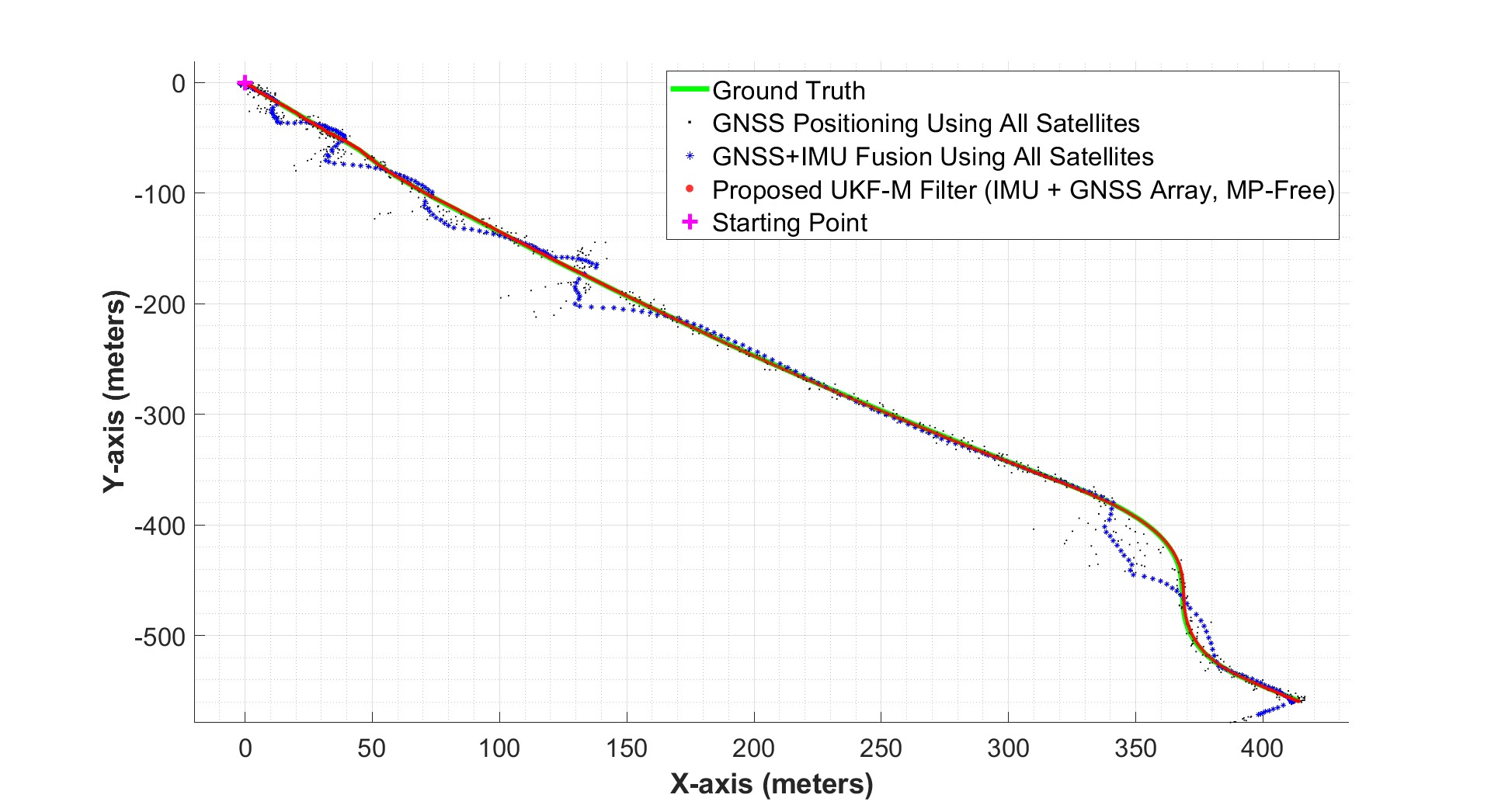}
        \caption{Suburban environment: Position MSE reduced from 6.7\,m to 0.56\,m, Attitude MAE = $0.15^\circ$, Success rate = 94\%.}
        \label{fig:suburban}
    \end{subfigure}
    \hfill
    \begin{subfigure}[b]{0.48\linewidth}
        \centering
        \includegraphics[width=\linewidth]{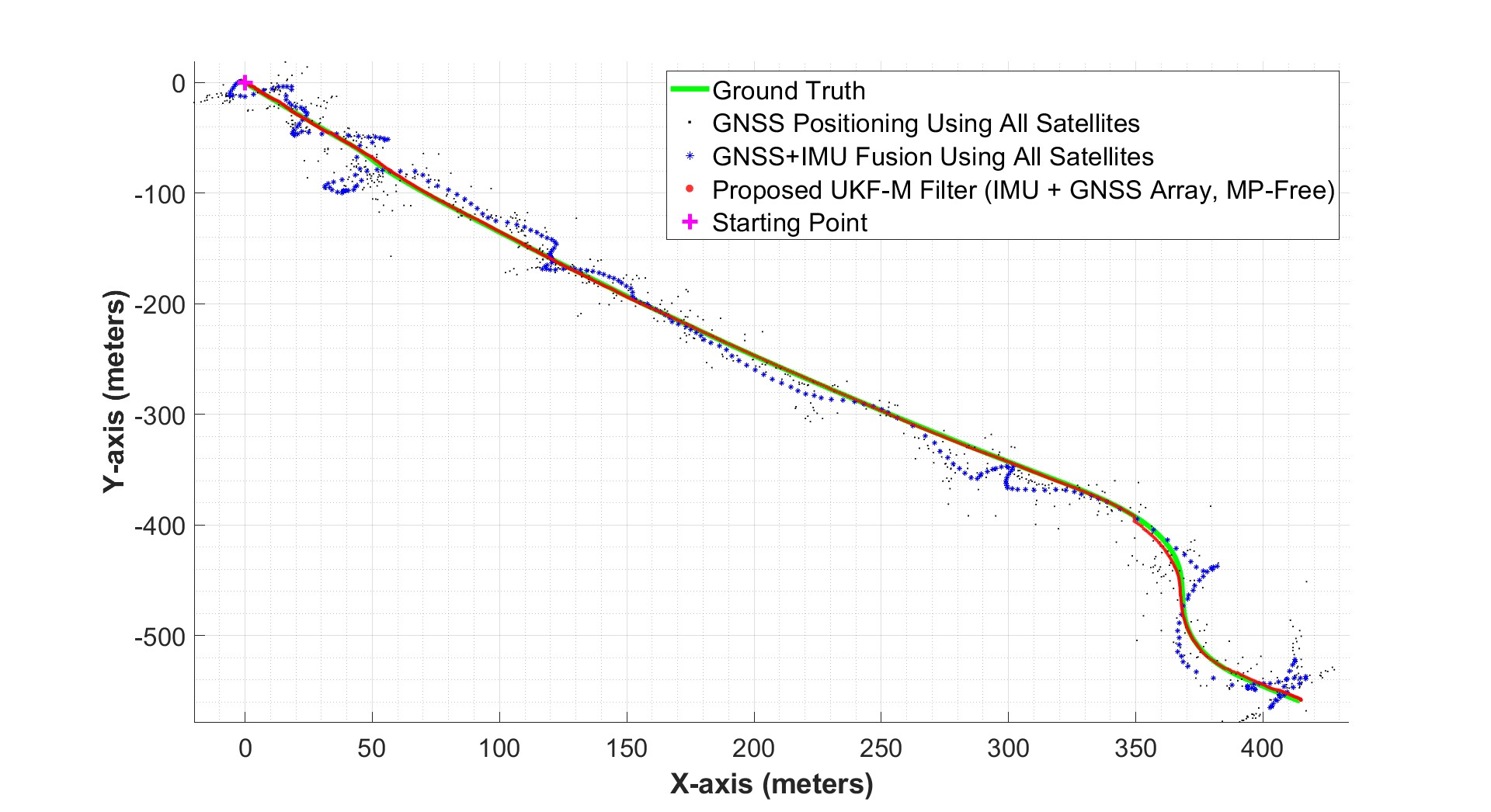} 
        \caption{Urban environment: Position MSE reduced from 23\,m to 1.7\,m, Attitude MAE = $0.47^\circ$, Success rate = 87\%.}
        \label{fig:urban}
    \end{subfigure}
    \caption{Comparison of filter performance in different environments. Green: Ground Truth; Black: GNSS positioning using all satellites; Blue: GNSS+IMU fusion using all satellites; Red: Proposed filter in Fig.~(\ref{fig:UKF_Block}).}
    \label{fig:traj_results}
\end{figure*}

\begin{figure*}[ht]  
    \centering  
    \includegraphics[width=0.9\linewidth]
    {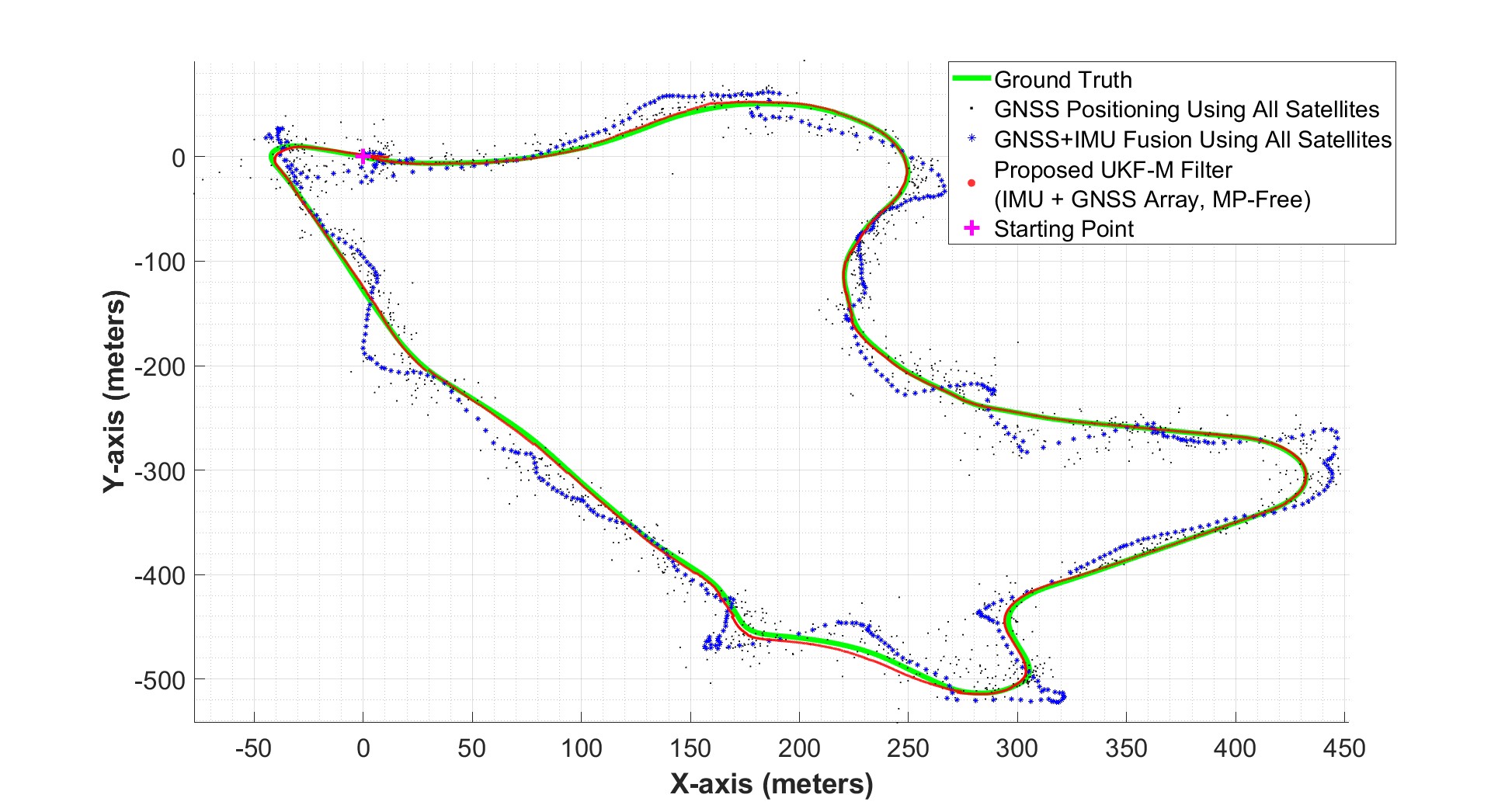}  
    \caption{Urban environment: Position MSE reduced from 34m to 2.7m, Attitude MAE = $1.47^\circ$, Success rate = 83\%.}  
    \label{fig:urban2}  
\end{figure*}

\subsection{Simulation 2: Moving Vehicle Scenario}
The second simulation models a realistic moving vehicle scenario using only the GPS constellation. Satellite positions are obtained from an almanac, and a \ang{15} elevation cutoff yields 10 visible satellites at the simulated KITTI trajectory location. Ground truth positions and IMU data are sourced from the KITTI dataset~\cite{geiger2013vision}, providing realistic motion and noisy inertial measurements. Multipath is modeled using urban parameters based on ITU recommendation for building distributions~\cite{series2013propagation}, with Rayleigh-distributed building heights. This setup enables testing across different environments (e.g., suburban vs. urban) by adjusting street widths and building heights. GNSS observations are modeled using the same method as in the static case, with multipath simulated per Eqs.~\eqref{eq:carrier_phase_error}--\eqref{eq:contaminated_pseudorange}.

Each satellite signal is evaluated for multipath contamination based on elevation angle and building geometry. if the surrounding building heights allow for a
reflection point that satisfies the law of reflection and enables the reflected signal to reach the antenna array, the signal is considered contaminated. The associated excess path length introduces pseudorange and carrier phase errors, with additive white noise applied to all signals.

Using simulated GNSS data and realistic IMU readings, the proposed filter in Fig.~(\ref{fig:UKF_Block}) integrates measurements from all antennas and the IMU to detect multipath (via Algorithm~\ref{alg:multipath_detection}) and improve positioning by using only clean signals. Figure~(\ref{fig:traj_results}) shows results for a KITTI trajectory in two simulated environments: one representing a suburban area and the other an urban setting with severe multipath interference. These results highlight the effectiveness of the proposed filter and detection algorithm in enhancing positioning accuracy. The filter successfully eliminates contaminated signals, significantly improving accuracy. The detection success rate and baseline MAE are also provided.

Fig.~(\ref{fig:urban2}) illustrates results for another KITTI trajectory in a simulated urban environment with strong multipath interference. These figures demonstrate the significant error in positioning due to severe multipath, with IMU fusion failing to mitigate the effect. However, by applying our proposed algorithm to exclude multipath-affected satellites and maintain attitude tracking for subsequent detections, more accurate positioning is achieved.

\section{Conclusion and Future Work} \label{Concl}
This study presents a GNSS array-based multipath detection algorithm integrating real-time attitude estimation using an Unscented Kalman Filter (UKF) on a manifold. By fusing GNSS observations and IMU data, the method improves both positioning accuracy and attitude estimation while leveraging a RANSAC-based algorithm to discriminate multipath-affected satellites, enhancing computational efficiency and reducing processing complexity.

We performed two simulations: the first tested the performance of the RANSAC-based algorithm in multipath detection, which resulted in a slightly lower success rate due to its overprotective behavior in excluding significantly noisy satellite signals but provided better attitude estimations. The second simulation confirmed the effectiveness of the proposed UKF-on-manifold-based method in urban environments, demonstrating significant reductions in position MSE and improvements in attitude estimation. The approach remains reliable as long as at least four satellites are multipath-free, ensuring robust performance in various conditions. However, since attitude estimates serve as references in subsequent epochs, any errors in estimation may propagate over time, potentially degrading overall system performance.

Future work will focus on experimental validation using physical GNSS receivers in real-world environments to assess practical applicability. Additional efforts will be directed toward refining the method to mitigate error propagation effects, ensuring stable long-term performance. Incorporating multi-frequency GNSS signals and integrating advanced positioning methods such as factor graph optimization will further enhance robustness and accuracy, making the approach more resilient in challenging urban environments.

\section*{Acknowledgments}
The authors gratefully acknowledge the support from the SDAIA-KAUST Center of Excellence on AI.

\vspace{0pt}

\bibliographystyle{IEEEtran}
\bibliography{ref}
\end{document}